\documentclass[conference]{IEEEtran}
\usepackage{graphicx}
\usepackage{times}
\usepackage{float}
\usepackage{eqnarray,amsmath}
\usepackage{array}
\usepackage{booktabs}
\usepackage{textcomp}
\usepackage{amsmath,amssymb,amsthm}
\usepackage{mathtools}
\usepackage{epstopdf}
\usepackage{multirow}
\usepackage{multicol}
\usepackage{pbox}
\usepackage{cite}
\usepackage{tikz}
\usepackage{algorithm,algpseudocode}
\usepackage{pifont}
\usepackage{breqn}
\usepackage{xcolor}
\usepackage[caption=false, font=footnotesize]{subfig}

\allowdisplaybreaks

\algnewcommand{\Inputs}[1]{%
  \State \textbf{Inputs:}\hspace*{\algorithmicindent}\parbox[t]{.8\linewidth}{\raggedright #1}
}
\algnewcommand{\Outputs}[1]{%
  \State \textbf{Outputs:}\hspace*{\algorithmicindent}\parbox[t]{.8\linewidth}{\raggedright #1}
}
\algnewcommand{\Initialize}[1]{%
  \State \textbf{Initialize:}\hspace*{\algorithmicindent}\parbox[t]{.8\linewidth}{\raggedright #1}
}

\makeatletter
\def\endthebibliography{%
  \def\@noitemerr{\@latex@warning{Empty `thebibliography' environment}}%
  \endlist
}
\makeatother

\def\NoNumber#1{{\def\alglinenumber##1{}\State #1}\addtocounter{ALG@line}{-1}}

\bstctlcite{IEEEexample:BSTcontrol}
\begin{document}
\title{Geometry based Stochastic Channel Modeling \\
using Ambit Processes }

\author{\IEEEauthorblockN{Rakesh R. T. and Emanuele Viterbo} \IEEEauthorblockA
{Department of Electrical and Computer Systems Engineering, Monash University, Clayton, VIC 3800, Australia\\
Email: \{rakesh.thankamani, emanuele.viterbo\}@monash.edu}}

\maketitle

\providecommand{\keywords}[1]{\textbf{\textit{Index terms---}} #1}

\begin{abstract}
The simulation of vehicular wireless channels using geometry-based radio channel models is computationally intensive when the number of scatterers is significantly high. In this paper, we propose a new geometry-based stochastic channel model to simulate and analyze the aforementioned channels based on a framework developed from the theory of \textit{ambit processes}.
Under reasonable assumptions, the underlying mathematical structure of the proposed channel model enables the characterization of high mobility channels in terms of fading statistics, spatio-temporal channel correlation, and Doppler spectrum, besides ensuring tractable analysis. The developed algorithm facilitates fast simulation of high mobility channels and accounts for key features of vehicular channels including appearance and disappearance of multi-path components, spatial consistency, and captures the correlation between time-evolving delay and Doppler associated with multi-path components. Finally, we carry out simulations to obtain crucial insights about the characteristics of typical vehicle-to-infrastructure channels based on the proposed channel model.
\end{abstract}

\IEEEpeerreviewmaketitle

\begin{keywords}
Geometry based stochastic channel model, ambit process, Le\'{v}y basis, birth-death process, temporal autocorrelation.
\end{keywords}

\section{Introduction}
\label{Introduction}
Vehicle-to-vehicle (V2V) and vehicle-to-infrastructure (V2I) communication systems form crucial components of  intelligent transportation systems \cite{wang2014cellular,Agiwal2016next}. In order to support the development and deployment of V2V and V2I communication systems, channel models characterizing important features of vehicular wireless channels are required. In recent years, several channel models  based on different modeling techniques have been proposed \cite{czink2009time,raschkowski2015metis,wu2015nonstationary}. These channel models offer varying degrees of flexibility in terms of simplicity, accuracy, and scalability to enable modeling of various channel parameters.

In order to balance the trade-off between accuracy and complexity of the channel model, geometry based stochastic channel model (GBSCM)s have found wide acceptance for modeling vehicular channels \cite{karedal2009geometry,wu2015nonstationary,dahech2017nonstationary}. In GBSCMs, scatterers are placed in the Euclidean plane based on probability density functions (PDF)s which are derived empirically from measurements, and the channel impulse response (CIR) is evaluated using simplified ray tracing techniques. Depending on the modeling aspects, GBSCMs further classified into regular-GBSCMs and irregular-GBSCMs. In regular-GBSCMs \cite{patzold2012mobile}, scatterers are distributed across certain regular shapes including rings and ellipses, whereas in irregular-GBSCMs, scatters are distributed across arbitrary shapes. Despite the simplicity and analytical tractability of regular-GBSCMs, they do suffer from inaccuracies in modeling vehicular channels. In general, placement of scatterers on the regular shapes does not always ensure modeling of real-world scenarios \cite{karedal2009geometry}. On the other hand, owing to the more realistic distribution of scatterers, irregular-GBSCMs show a better agreement with the measurement data \cite{wang2018survey}.

Despite the accurate modeling of vehicular channels, irregular-GBSCMs tend to suffer from high simulation complexity and reduced analytical tractability due to the involvement of a large number of scatterers used to characterize the multi-path environment. Due to the high computational complexity resulting from the involvement of a large number of scatterers, existing methods traditionally employ approximations to model scattering environments. One such approach \cite{zajic2006efficient,patzold2012mobile} relies on specific placement of a finite number of scatterers in the scattering region so that the approximated model captures channel characteristics with a fair amount of accuracy. However, this approach fails to achieve desirable statistical properties of the reference model \cite{Kaltenberger2007low}. The second approach \cite{czink2010low} involves the introduction of discrete prolate spheroidal sequences to represent scattering environments thereby reducing the dimensionality of the problem. The approach in \cite{czink2010low} is based on a key assumption that wireless channels are wide-sense stationary, and variations in direction of arrival or departure as well as multi-path power during the simulation interval are negligible. However, in realistic propagation scenarios, the vehicular wireless channel is, in general, non-stationary \cite{karedal2009geometry}, and therefore, channel simulations for longer time duration mandate multi-path powers, path delays, and  Doppler frequencies to be calculated in every channel instance using direct geometrical calculations. Moreover, the approach in \cite{czink2010low} does not model the birth-death event of multi-path components.

To address the computational complexity, in this paper, we present an irregular GBSCM in which we model scatterer locations and their modification relative to a moving wireless node using a novel approach known as \textit{ambit process.} We have two main objectives: (i) To develop a general analytical framework for irregular-GBSCM characterizing spatial consistency, non-stationarity, birth-death events of multi-path components, (ii) Simulate channel variations resulting from change in direction of multi-path components and their appearance/disappearance in a computationally efficient way.

The rest of the paper is organized as follows. Section \ref{An_Approach_to_Geometry_Based_Channel_Modeling} presents the system model as the basis for the proposed geometrical channel modeling and the fundamentals of ambit processes. Section \ref{A_Simulator_for_V2I_Channels_based_on_the_Proposed_Channel_Model} elaborates a novel 2D convolution based algorithm for simulating V2I channels. Section \ref{Numerical_Simulations} includes simulation results and Section \ref{Conclusion} concludes the paper.
%%%%%%%%%%%%%%%%%%%%%%%%%%%%%%%%%%%%%%%%%%%%%%%%%%%%%%%
\section{An Approach to Geometry Based Channel Modeling}
\label{An_Approach_to_Geometry_Based_Channel_Modeling}
\subsection{Theory of Ambit Processes}
\label{Theory_of_Ambit_Processes}
\begin{figure}[H]
	\centering
	%crop order 'left''bottom''right''top'
	\includegraphics[trim=9.2cm 4.5cm 3.2cm 4.8cm, clip=true, totalheight=0.21\textheight]{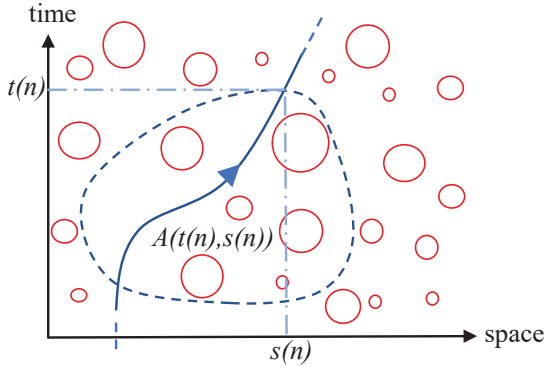}
	\caption{Depiction of an ambit process.}	
	\label{fig:Ambit_Proc}
\end{figure}

In this section, we provide the theoretical foundation of ambit processes which is subsequently leveraged to model vehicular wireless channels. Ambit processes are generalized stochastic processes which describe dynamic phenomena in time and space \cite{barndorff2018ambit}. The basic settings for an ambit process includes a stochastic field $X(t,\textbf{s})$ in time-space $\mathbb{R}_{+}\times\mathbb{R}^{k}$ and a curve $\mathcal{C}(n)=(t(n),\textbf{s}(n))$ with the sampling point, $n\in \mathbb{N}$ \cite{barndorff2018ambit}. As shown in Fig.~\ref{fig:Ambit_Proc}, the curve $\mathcal{C}$ presents the trajectory of a moving object in the time-space plane with each point in the curve (indexed by $n\in\mathbb{N}$) representing the location in space of the object at a given instance of time $t(n)$. The stochastic field is the random phenomena experienced by the object at the time-space sampling point $\mathcal{C}(n)$. The stochastic field at the point $\mathcal{C}(n)$ can be influenced by the randomness (modeled through L\`{e}vy bases \cite{barndorff2018ambit}) observed at some time-space point.
In the context of vehicular wireless channels, the curve $\mathcal{C}$ can be used to specify the trajectory for a mobile user (MU) and the stochastic field $X$ can be considered as the channel gain experienced by the MU. The coupling between the randomness occurs at these points and the stochastic field experienced on the curve is generally modeled by a combination of deterministic functions and random variables (known as {\em volatility} terms). The circles depicted in Fig.~\ref{fig:Ambit_Proc} relate to the {\em volatility} or {\em additional randomness} associated to a point in the time-space plane \cite{barndorff2018ambit}. It may be noted that not all the time-space points can influence the stochastic field at a sampling point $\mathcal{C}(n)$. The stochastic field at any sampling point $n\in\mathbb{N}$ on the curve depends only on the time-space randomness occurring prior to the time, $t(n)$. The collection of these time-space points is called an {\em ambit set} of $(t(n),\mathbf{s}(n))$ which is represented by the subset $A(t(n),\mathbf{s}(n))$ of $\mathbb{R}_{+}\times\mathbb{R}^{k}$.

An example of ambit set is shown in Fig.~\ref{fig:Ambit_Proc} in terms of a closed region with boundaries represented by the dotted curve. In general, the stochastic field $X(t,\textbf{s})$ at an arbitrary point ($t$, $\textbf{s}$) in the time-space plane is represented by the combination of a constant term and stochastic components. The stochastic components are represented either through deterministic or stochastic integrals. From a mathematical perspective, the stochastic field $X(t,\textbf{s})$ is represented by,
\begin{align}\label{eq_ambit_field}
    X(t,\textbf{s})=&\mu+\int_{B(t,\textbf{s})}g(t,\textbf{s};u,\textbf{p})\eta(u,\textbf{p})\text{d}u\text{d}\textbf{p}\nonumber\\&+\int_{A(t,\textbf{s})}f(t,\textbf{s};u,\textbf{p})\sigma(u,\textbf{p})L(\text{d}u,\text{d}\textbf{p}),
\end{align}
where $\mu$ is the {\em drift level}, $A(t,\textbf{s})$ and $B(t,\textbf{s})$ represent ambit sets, $f$ and $g$ are deterministic functions known as \textit{kernel functions}.  The variables, $\sigma$ and $\eta$ denote stochastic fields referred to {\em volatility} at a given point from the ambit sets, $A(t,\mathbf{s})$ and $B(t,\mathbf{s})$, respectively. The dynamics associated with the time-space points is primarily captured via the L\`{e}vy basis $L$. Typical examples of L\`{e}vy bases are Gaussian, Poisson, $\alpha-$stable, and generalized hyperbolic distributions. As an example, in the subsequent sections, we apply the theory of ambit process to model V2I channels characterized by static scatterers.
\subsection{Characterization of V2I Channels}
\label{Characterization_of_V2I_Channels}
\begin{figure}[H]
	\centering
	%crop order 'left''bottom''right''top'
	\includegraphics[trim=6.8cm 3.4cm 3.4cm 1.9cm, clip=true, totalheight=0.18\textheight]{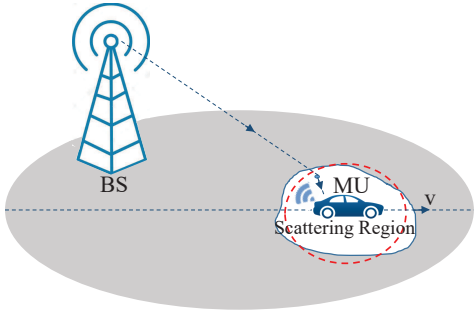}
	\caption{A scenario depicting BS-MU channel.}	
	\label{fig_BS_MU_channel}
\end{figure}

Consider a base station (BS) to MU channel. We assume that the MU travels in a prescribed direction with a time-varying velocity $v(t)$ (with $v_{0}=v(t=0)$) as shown in Fig.~\ref{fig_BS_MU_channel}. The distance between the BS and MU at the time instance, $t$, is denoted by $D(t)$. Scatterers which cause multi-path signal propagation are assumed to be static. To model the scattering environment, we consider a macro-cellular environment with local scattering effect where scatterers are generally present only inside a region around MU, as shown in Fig.~\ref{fig_BS_MU_channel}. We assume that $D(t)$ is sufficiently large for a 2D scattering region to model the 3D V2I channel. The shaded region shown in Fig.~\ref{fig_BS_MU_channel} represents the location of effective scatters in the BS-MU channel scenario. The region of scatterers can be approximated by a circular disc around the MU as shown in Fig.~\ref{fig_BS_MU_channel}. The scatterers visible to the MU are essentially located within the radius of the disc. A similar approximation for  scatterer distribution has found extensive use in existing literature \cite{Ertel1999angle}. The radius of the circular disc, $R$, is generally chosen based on the minimum signal power the MU can detect. With $N(t)$ scatterers present inside the circular scattering region at time instant, $t$, we model the time-variant complex impulse response of the channel as the superposition of $N(t)+1$ multi-paths by \cite{karedal2009geometry},
\begin{align}\label{ch_imp_resp}
    h(t,\tau)= & \; a_{L}(t)e^{jkd_{L}(t)}\delta(t-\tau_{L}(t))\nonumber\\&+\sum_{s=1}^{N(t)}a_{s}e^{jkd_{s}(t)}\delta(t-\tau_{s}(t)),
\end{align}
where the first term in (\ref{ch_imp_resp}) represents the multi-path component due to line-of-sight (LoS) signal propagation, and the second term represents the single bounce multi-path components due to scatterers present inside the circular disc. Approximating signal propagation mechanism with single bounce scattering is widely accepted for channel modeling and also shows good match with measurements \cite{karedal2009geometry}. The variables $a_{L}(t)$ and $a_{s}(t)$ denote time-varying complex channel amplitudes due to LoS component and the single bounce components from $s-$th scatterer ($s\in\{1,2,...,N(t)\}$), respectively. The parameter, $k$ represents the {\em wave number}, which is related to the wavelength, $\lambda$, by the relation, $k=2\pi\lambda^{-1}$. $d_{L}(t)$ and $d_{s}(t)$ denotes time-varying signal propagation distance for LoS and $s-$th scattering multi-path component, respectively. The signal propagation delays are denoted by $\tau_{L}(t)=d_{L}(t)c^{-1}$ and $\tau_{s}(t)=d_{s}(t)c^{-1}$, respectively. The parameter, $c$ denotes velocity of the light in air. Similar to the approach in \cite{karedal2009geometry}, the complex amplitude of individual multi-path components can be expressed as,
\begin{align}
    a_{L}(t)=G_{L}^{1/2}d_{L}(t)^{-\gamma/2}
\end{align}
\begin{align}
    a_{s}(t)=G_{s}^{1/2}c_{s}(t)d_{s}(t)^{-\gamma/2},
\end{align}
where $G_{L}$ and $G_{s}$ are reference power terms which can be determined from channel measurements, $\gamma$ denotes path loss exponent, and $c_{s}(t)$ is a zero mean Gaussian random process chosen in order to maintain consistency between the model and measurement data. In the proposed channel modeling framework, we model time-varying channels based on the theory of ambit processes. For this purpose, we use an ambit process described in the time-space plane with dimension $\mathbb{R}_{+}\times \mathbb{R}$. The scatterers are treated as points and therefore, the time-space distribution of scatterers is modeled using a Poisson L\'{e}vy basis with a non-uniform intensity function, $\lambda_{s}(z)$, $\forall z\in A(t,\mathbf{s})$. A L\'{e}vy basis $L(A)$ is said to be a Poisson L\'{e}vy basis if $L(A)\sim \text{Pois}(\int_{A}\lambda_{s}(z)\text{d}z)$, $\forall A\in\mathcal{B}(\mathbb{R}_{+}\times \mathbb{R})$, where Pois$(\cdot)$ denotes poisson random variable and $\mathcal{B}$ denotes Borel set defined on the space $\mathbb{R}_{+}\times \mathbb{R}$.

We first establish the geometrical relationship between the framework of ambit processes discussed in Section \ref{Theory_of_Ambit_Processes} and the deployment scenario considered in Fig.~\ref{fig_BS_MU_channel}. Without loss of generality, we can choose the $x-$axis to be in parallel with the direction of motion, whereas the space dimension is taken perpendicular to the $x-$axis. As such, a mapping from time-axis in the ambit framework to $x-$axis is performed based on the velocity of motion of MU. The time-varying channel impulse response is modeled as the time-space varying stochastic ambit field experienced at the MU, i.e. $h(t,\tau)$ in (\ref{ch_imp_resp}) can be reformulated as,
\begin{align}\label{eq_appl_ambit}
    h(t,\tau)=&h\left(t,y_{i},u,y_{j}:(u,y_{j})\in A(t,y_{i})\right)\nonumber\\=&a_{L}(t)e^{jkd_{L}(t)}\delta(t-\tau_{L}(t))\nonumber\\&+\int_{A(t,y_{i})}f(t,y_{i};u,y_{j})\sigma(u,y_{j})L(\text{d}u,\text{d}y_{j}).
\end{align}
The variable $\tau$ denotes multi-path propagation delay. We model instantaneous $\tau$ as a function of time-space points inside the time-varying ambit set $A(t,y_{i})$, which models instantaneous scattering region around MU. The coordinates, $(t,y_{i})$ denote the instantaneous location of MU in the time-space plane. The power distribution of individual multi-path components is carefully modeled by choosing appropriate kernel function, $f$, and stochastic volatility function, $\sigma$. Without loss of generality, we assume the initial location of the MU to be at $(0,y_{i})$, where $y_{i}$ is chosen as a constant. To make the model consistent with the one defined in (\ref{ch_imp_resp}), we define the kernel function as,
\begin{equation}\label{kernal_func}
    f(t,y_{i};u,y_{j})=G_{s}^{1/2}d_{s}(t)^{-\gamma/2}e^{jkd_{s}(t)}\delta(t-\tau_{s}(t)),
\end{equation}
where,
\begin{align}\label{prop_dist}
d_{s}(t)=&d\left(BS_{x},BS_{y};\int_{0}^{u}\hspace{-0.25cm}v(\zeta)\text{d}\zeta,y_{s}\right)\nonumber\\&~~~~~~~~~~~~~~~+d\left(\int_{0}^{t}\hspace{-0.25cm}v(\zeta)\text{d}\zeta,y_{i};\int_{0}^{u}\hspace{-0.25cm}v(\zeta)\text{d}\zeta,y_{s}\right)
\end{align}
where $d(x_{1},y_{1};x_{2},y_{2})=\sqrt{(x_{1}-x_{2})^{2}+(y_{1}-y_{2})^{2}}$ is the Euclidean distance between the points, $(x_{1},y_{1})$ and $(x_{2},y_{2})$. The first and second terms in (\ref{kernal_func}) are the BS-to-scatterer distance and scatterer-to-MU distance, respectively. $BS_{x}$ and $BS_{y}$ are the $x$ and $y$ coordinates of BS location. Moreover, we select the volatility function, $\sigma(u,y_{j})=c_{s}(u,y_{j})$.
\subsection{Choosing Parameters for the Proposed V2I Channel Model}
\label{Choosing_Parameters_for_the_Proposed_V2I_Channel_Model}
Parameters of the proposed channel model can readily be selected based on reported measurements. The density of scatterers, $\lambda_{s}$ and the size of the ambit set, $R$ can be determined as follows. For simplicity, let us assume the underlying L\`{e}vy process defining the ambit process is homogeneous, i.e., $\lambda_{s}(z)=\lambda_{s}$, $\forall z\in \mathbb{R}_{+}\times\mathbb{R}$ and $v(t)=v_{0}$ within the simulation duration. Consider $N_{s}$ and $R_{s}$ be the average number of resolvable multi-path components and the average arrival rate of multi-path components, respectively. Then, the radius of the circular scattering region $R$ and the density parameter $\lambda_{s}$ can be evaluated as,
\begin{equation}\label{radius_ambit_set}
    R=\sqrt{\frac{N_{s}}{\lambda_{s}\pi}},
\end{equation}
\begin{equation}\label{density_scatterer}
    \lambda_{s}=\frac{R_{s}}{2Rv_{0}}
\end{equation}
Without loss of generality, parameters for V2I channels characterized by non-uniform spatial distribution of scatterers can be modeled using inhomogenous or non-stationary ambit processes. The model parameters can be determined by applying suitable modifications to (\ref{radius_ambit_set}) and (\ref{density_scatterer}).
%%%%%%%%%%%%%%%%%%%%%%%%%%%%%%%%%%%%%%%%%%%%%%%%%%%%
\section{A Simulator for V2I Channels based on the Proposed Channel Model}
\label{A_Simulator_for_V2I_Channels_based_on_the_Proposed_Channel_Model}
In this section, we present an algorithm for simulating time-varying CIR of non-stationary V2I channels. The computational complexity of the algorithm is significantly reduced through a 2D convolution operation. For this purpose, we transform the kernel function, $f$ in (\ref{kernal_func}) in the form of a semi-stationary function \cite{barndorff2018ambit}. Therefore, we decompose $f$ in (\ref{kernal_func}) into the format given in (\ref{kernel_func_modif}).
\begin{figure*}
\begin{align}\label{kernel_func_modif}
    f(t,y_{i};u,y_{j})\approx &\; G_{s}^{1/2}d\Big(BS_{x},BS_{y};\int_{0}^{u}\hspace{-0.15cm}v(\zeta)\text{d}\zeta,y_{s}\Big)^{-\gamma/2}\hspace{-0.7cm}\exp\Big(jk\Big(d\Big(BS_{x},BS_{y};\int_{0}^{u}\hspace{-0.15cm}v(\zeta)\text{d}\zeta,y_{s}\Big)+d\Big(\int_{0}^{t}\hspace{-0.15cm}v(\zeta)\text{d}\zeta,y_{i};\int_{0}^{u}\hspace{-0.15cm}v(\zeta)\text{d}\zeta,y_{s}\Big)\Big)\Big)\nonumber\\&\times\delta\Big(t-c^{-1}\Big(d\Big(BS_{x},BS_{y};\int_{0}^{u}\hspace{-0.15cm}v(\zeta)\text{d}\zeta,y_{s}\Big)+d\Big(\int_{0}^{t}\hspace{-0.15cm}v(\zeta)\text{d}\zeta,y_{i};\int_{0}^{u}\hspace{-0.15cm}v(\zeta)\text{d}\zeta,y_{s}\Big)\Big)\Big)
\end{align}
\hrule
\end{figure*}
The approximation, $d_{s}(t)\approx d(BS_{x},BS_{y};\int_{0}^{u}\hspace{-0.15cm}v(\zeta)\text{d}\zeta,y_{s})$ is expected to be sufficiently accurate since $d(BS_{x},BS_{y};\int_{0}^{u}\hspace{-0.15cm}v(\zeta)\text{d}\zeta,y_{s}) \gg d(\int_{0}^{t}\hspace{-0.15cm}v(\zeta)\text{d}\zeta,y_{i};\int_{0}^{u}\hspace{-0.15cm}v(\zeta)\text{d}\zeta,y_{s})$ in general. For realization, we initially assume constant velocity $v_{0}$ for $v(t)$ and introduce time variability later.
By treating the terms $\exp(kd(v_{0}t,y_{i};v_{0}u,y_{j}))$ and $\exp(kd(BS_{x},BS_{y};v_{0}u,y_{j}))$ as constants for a given time instance $t$, the kernel function in (\ref{kernel_func_modif}) is represented in the form,
\begin{equation}\label{kernel_func_final}
    f(t,y_{i};u,y_{j})=\overline{f}(t-u,y_{i}-y_{j})*\widehat{f}(t,u,y_{j}),
\end{equation}
where 
\begin{equation}
    \overline{f}(t-u,y_{i}-y_{j})=e^{jkd(v_{0}t,y_{i};v_{0}u,y_{j})}\delta\left(t-\frac{d(v_{0}t,y_{i};v_{0}u,y_{j})}{c}\right)
\end{equation}
\begin{align}
    \widehat{f}(u,y_{j})=&G_{s}^{1/2}e^{jkd(BS_{x},BS_{y};v_{0}u,y_{j})}\nonumber\\&\times\delta\left(t-\frac{d(BS_{x},BS_{y};v_{0}u,y_{j})}{c}\right).
\end{align}
The symbol $*$ denotes linear convolution operator. Now, we apply a convolution technique to evaluate the time-varying channel impulse response by extending the method introduced in \cite{bennedsen2014discretization}, which was originally proposed for simulating one dimensional L\`{e}vy processes. We first modify the expression in (\ref{eq_appl_ambit}) by separating time and space integration. For this purpose, we assume the ambit set has the form, $A(t,y_{i})=(t,0)+A$, where $A=A(0,y_{i})$. Hence,
\begin{align}\label{eq_appl_ambit_modif}
    h(t,\tau)=&\int_{0}^{\infty}\int_{A}\left[\overline{f}(t-u,y_{i}-y_{j})*\widehat{f}(u,y_{j})\right]\nonumber\\&
    ~~~~~~~~~~\times\sigma(u,y_{j})L(\text{d}u,\text{d}y).
\end{align}
We assume that $\overline{f}(x,y_{i}-y_{j})=0$ for $x\leq 0$.
Moreover, based on the independent scattering property of L\`{e}vy bases \cite{barndorff2018ambit}, we write, $L(\text{d}u,\text{d}y)=\text{d}L(u,y_{i})$. Therefore, (\ref{eq_appl_ambit_modif}) becomes
\begin{align}\label{eq_appl_ambit_modif2}
    h(t,\tau)=&\int_{0}^{\infty}\int_{A}\overline{f}(t-u,y_{i}-y_{j})*\Big[\widehat{f}(u,y_{j})\nonumber\\&
    ~~~~~~~~~~\times\sigma(u,y_{j})\text{d}L(u,y_{j})\Big].
\end{align}
We see that the resulting ambit process can be represented in terms of a $2$D grid with step sizes, $\Delta>0$ and $\Lambda>0$. Therefore, we have,
\begin{align}
    h(k\Delta,\tau)=&\sum_{l=-\infty}^{\infty}\sum_{m=-\infty}^{\infty}\int_{(l-1)\Delta}^{l\Delta}\int_{(m-1)\Lambda}^{m\Lambda}\overline{f}(k\Delta-u,y_{i}-y_{j})\nonumber\\&
    *\Big[\widehat{f}(u,y_{j})\sigma(u,l\Lambda)\text{d}L(u,y_{j})\Big].
\end{align}
We note that $\overline{f}(x,y)=0$, if $(x,y)\notin A(0,y_{i})$. 
Similar to the approach in \cite{bennedsen2014discretization}, we assume $\overline{f}$, $\widehat{f}$, and $\sigma$ are approximately constant and equal to their respective values at $((l-1)\Delta,(m-1)\Lambda)$. Then,
\begin{align}\label{approx_imp_resp}
    h(k\Delta,\tau)\approx&\sum_{l=-\infty}^{\infty}\sum_{m=-\infty}^{\infty}\overline{f}((k-l+1)\Delta,y_{i}-(m-1)\Lambda)\nonumber\\&
    *\Big[\widehat{f}((l-1)\Delta,(m-1)\Lambda)\nonumber\\&\times\sigma((l-1)\Delta,(m-1)\Lambda)\nabla L_{l,m}\Big],
\end{align}
where $\nabla L_{l,m}=L(l\Delta,m\Lambda)-L((l-1)\Delta,(m-1)\Lambda)$, $l,m\in \mathbb{N}$ are the increments of the underlying L\`{e}vy process. Now, we devise a simulation algorithm for V2I channel in Algorithm \ref{Non-stationary_V2I_channel_simulator} based on the formulation in (\ref{approx_imp_resp}). 
\begin{algorithm}[H]
		\caption{Non-stationary V2I channel simulator}
		\label{Non-stationary_V2I_channel_simulator}
		\begin{algorithmic}[1]
		\Inputs{Choose appropriate values of $\Delta$, $\Lambda$, $\tau_{max}$, $\delta \tau$, $T_{max}$;}
		\Initialize{define $M=\text{floor}(R/\Lambda)$, $N=\text{floor}(R/v\Delta)$, $P=\text{floor}(T_{max}/\Delta)$,  $D=\text{floor}(\tau_{max}/\delta\tau)$; $Z_{mat}=[0]_{(2(N+P)-1)\times (2D-1)}$; create position vector of MU $x(j\Delta)$ where $j\in\{1,2,..,N+P-1\}$; $x_{0}=0$}
		\For{each $i\in\{-M,-M+1,...,M-1\}$}
		\State Initialize $X_{mat}=[0]_{2N\times D}$ and $Y_{mat}=[0]_{2P\times D}$
		\State Compute $D_{1}=\text{floor}(c^{-1}d(j\Delta,0;i\Lambda,0)/\delta\tau)$ \NoNumber{for $j\in\{-N,-N+1,...,N-1\}$}
		\State $X_{mat}(j,D_{1})=\overline{f}(j\Delta,i\Lambda)$
		\State $D_{2}=\text{floor}(c^{-1}d(BS_{x},BS_{y};j\Delta,i\Lambda)/\delta\tau)$ \NoNumber{for $j\in\{0,1,...,P-1\}$}
		\State  $Y_{mat}(j,D_{2})=\widehat{f}(j\Delta,i\Lambda)\sigma(j\Delta,i\Lambda)\nabla _{j+1,i+1}$
		\State Compute 2D convolution of $Y_{mat}$ and row wise fold \NoNumber{$X_{mat}$ and equate the convoluted sequence to $Z_{mat,i}$}
		\State $Z_{mat}=Z_{mat}+Z_{mat,i}$
		\EndFor
		\For{each $i\in\{1,2,..,N+P-1\}$}
		\State Calculate the instantaneous velocity $v_{i}$ based on the \NoNumber{adopted model for time-varying velocity of MU}
		\State Determine the new location of MU $x_{i}=x_{i-1}+v_{i}\Delta$
		\State Identify the $j$ such that $x(j\Delta)\leq x_{i}\leq x((j+1)\Delta)$
		\State Compute $\alpha=[x_{i}-x(j\Delta)]/[x((j+1)\Delta)-x(j\Delta)]$
		\State $h_{N}(i\Delta,\tau)=\alpha Z_{mat}((j+1)\Delta,\tau)$\NoNumber{~~~~~~~~~~~~~~~~~~~~~~~~~~~~$+(1-\alpha) Z_{mat}(j\Delta,\tau)$}
		\EndFor
		\State \textbf{Output:} 2D matrix $h_{N}$
		\end{algorithmic}	
\end{algorithm}
The algorithm has two phases; initialization and channel simulation. During the initialization phase, all the required channel parameters in the form of matrices $X_{mat}$ and $Y_{mat}$ are precomputed for the 2D convolution operation to be used in the channel simulation phase. For this purpose, we first sample time, space ($y-$ dimension), and delay dimensions with respective number of points $P$, $M$, and $D$. The parameters $P$, $M$, and $D$ are determined based on maximum propagation delay $\tau_{max}$, delay resolution $\delta\tau$, simulation time $T_{max}$, and radius of the scattering region $R$. Then, for each space sample index, we construct two matrices with columns corresponding to delay and rows corresponding to time. For the first matrix $X_{mat}$, we compute the function $\overline{f}$ for all the time and space sample points. The computed $\overline{f}$ is mapped into the locations of $X_{mat}$ specified by the time and delay sample points. For the second matrix $Y_{mat}$, a similar procedure is adopted with the function defined by the product of $\widehat{f}$, $\sigma$, and $\nabla$. In the simulation phase, we perform 2D convolution of row wise folded $X_{mat}$ and $Y_{mat}$ for each space sample points as shown in step 8 of Algorithm. The resulting 2D matrices for each space sample index are added.

The computational benefit is obtained owing to the fact that the proposed V2I channel simulator tracks modifications in the received signal power, propagation delay, and phase of each multi-path component through the convolution operation. Computationally efficient algorithms exist for implementing convolution of sequences. It should be noted that impulse response for a longer time duration $T_{max}$ can be efficiently computed by breaking the matrix $Y_{mat}$ into multiple small matrices and apply overlap-add or overlap-save method to compute their convolution.

The effect of time-varying velocity of MU on channel impulse response is introduced in steps 11-16 of Algorithm. The time varying velocity of MU results in sampling of the MU's location at a non-uniform rate depending on the velocity of MU at a given instant of time. The impulse response of a non-stationary V2I channel $h_{N}(t,\tau)$ is simulated using a non-uniform sampling rate conversion technique given in steps 12-16 of the Algorithm. We assume that the actual space-sampling interval (along x-axis) for simulating V2I channel is sufficiently small so that adjacent samples of the channel coefficients are highly correlated. Therefore, we apply a linear interpolation technique to determine the channel coefficient at the required spatial point as given in step 15-16 of Algorithm. The non-stationary time-varying channel impulse response is stored in variable $h_{N}$.
%%%%%%%%%%%%%%%%%%%%%%%%%%%%%%%%%%%%%%%%%%%%%%%%%%%%
\section{Numerical Simulations \label{Numerical_Simulations} }
To validate the proposed channel modeling framework, we carry out numerical simulations. We consider that a BS is arbitrarily located at $(-100,20)$ and the initial position of a MU is at $(0,0)$. The MU is assumed to be moving away from  the BS with an initial velocity, $v_{0}=40$ km/hr as shown in Fig.~\ref{fig_BS_MU_channel}. We assume that BS transmits signal at a frequency $f_{c}=2.6$ GHz. Moreover, $G_{s}=G_{L}=(\lambda/4\pi)^{2}$ and path loss exponent, $\gamma=1.7$ (chosen based on the average of path loss exponents reported in \cite{karedal2009geometry}). Numerical results corresponding to the temporal autocorrelation function (ACF) and Doppler power spectral density (PSD) are generated by averaging the output of 500 independent realization of the Poisson L\`{e}vy field. Moreover, LoS component is not accounted in the simulations.
\vspace{-0.4cm}
\begin{figure}[H]
	\centering
	\includegraphics[height=6.7cm,width=8.7cm]{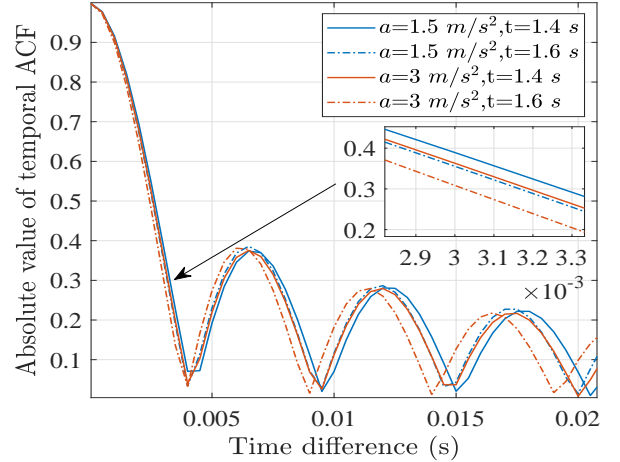}
	\caption{Temporal ACFs of the simulated V2I channel determined at different time instants for the accelerating MU.}
	\label{fig:Temporal_ACF_non_stationary}
\end{figure}
\vspace{-0.85cm}
\begin{figure}[H]
	\centering
	\includegraphics[height=6.7cm,width=8.7cm]{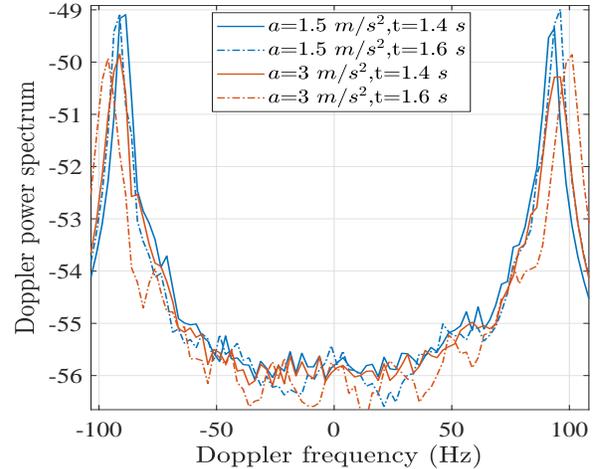}
	\caption{Doppler PSD of the simulated V2I channel determined at different time instants for the accelerating MU}
	\label{fig:Temporal_Doppler_non_stationary}
\end{figure}
We simulate non-stationary V2I channels by accounting time-varying velocity for MU.  We assume that the MU accelerates with a uniform rate of acceleration $a$. The simulation plots for normalized temporal ACF of the channel at different time instances are shown in Fig.~\ref{fig:Temporal_ACF_non_stationary}. The plots show that as the velocity of the MU increases with time due to the acceleration, the temporal ACF decreases rapidly with the time difference. This implies that channel decorrelates quickly with time. The time-varying Doppler PSD corresponding to the received signal at MU is presented in Fig.~\ref{fig:Temporal_Doppler_non_stationary}. The shape of the Doppler PSD closely follows the measured Doppler PSD for typical V2I channels.
The Doppler PSD evaluated at different time instances varies significantly due to the presence of non-stationary properties of the simulation environment. This is essentially due to the inclusion of time varying velocity for MU and the birth-death process on time-space axis. The plots in Fig.~\ref{fig:Temporal_Doppler_non_stationary} indicate that maximum Doppler shift increases for higher rate of accelerations since more acceleration leads to higher velocities thereby increasing the Doppler shift.
\vspace{-0.65cm}
\begin{figure}[H]
    \centering
    \subfloat[\label{1a}]{%
       \includegraphics[width=0.49\linewidth]{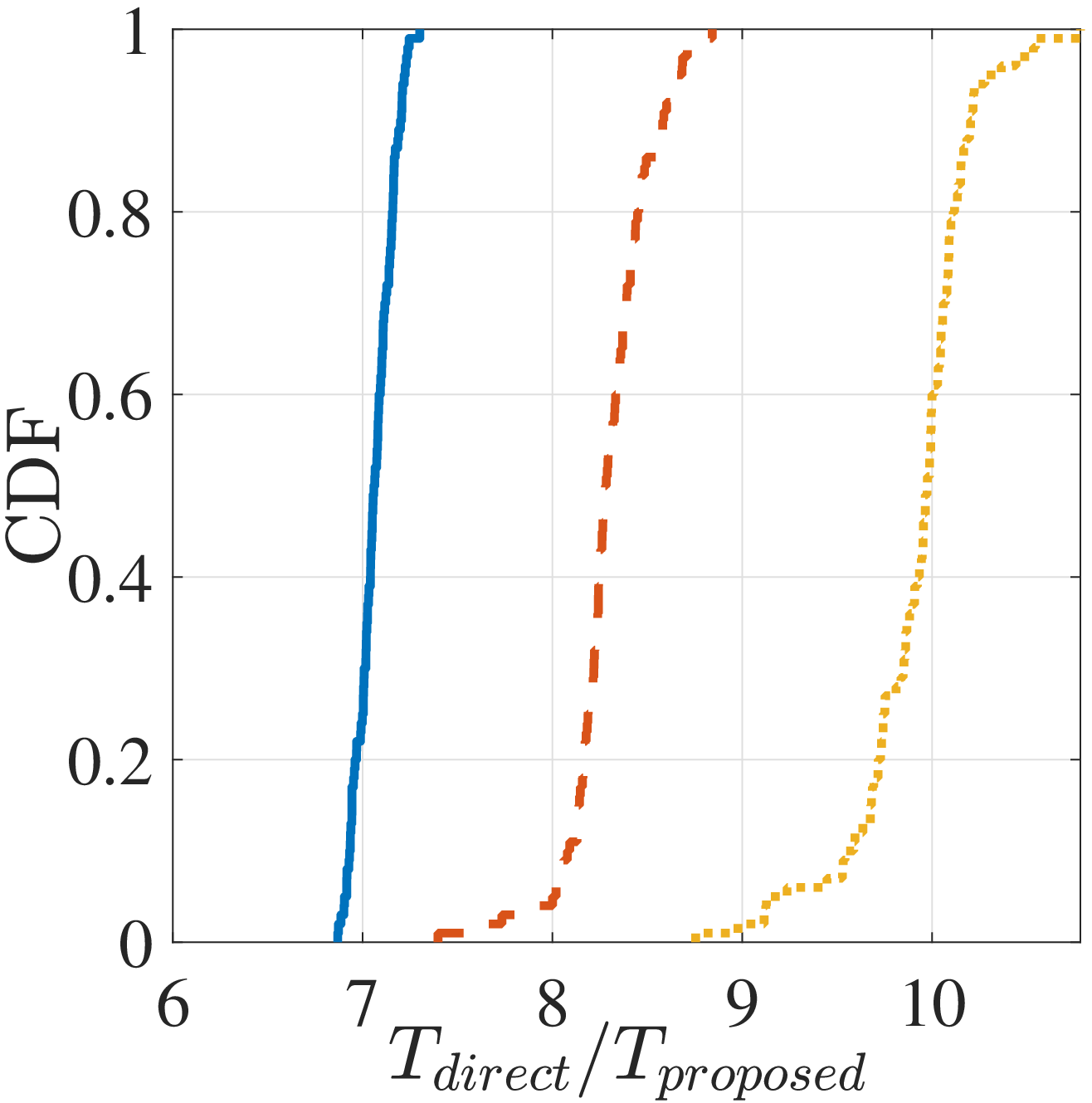}}
    \hfill
  \subfloat[\label{1b}]{%
        \includegraphics[width=0.49\linewidth]{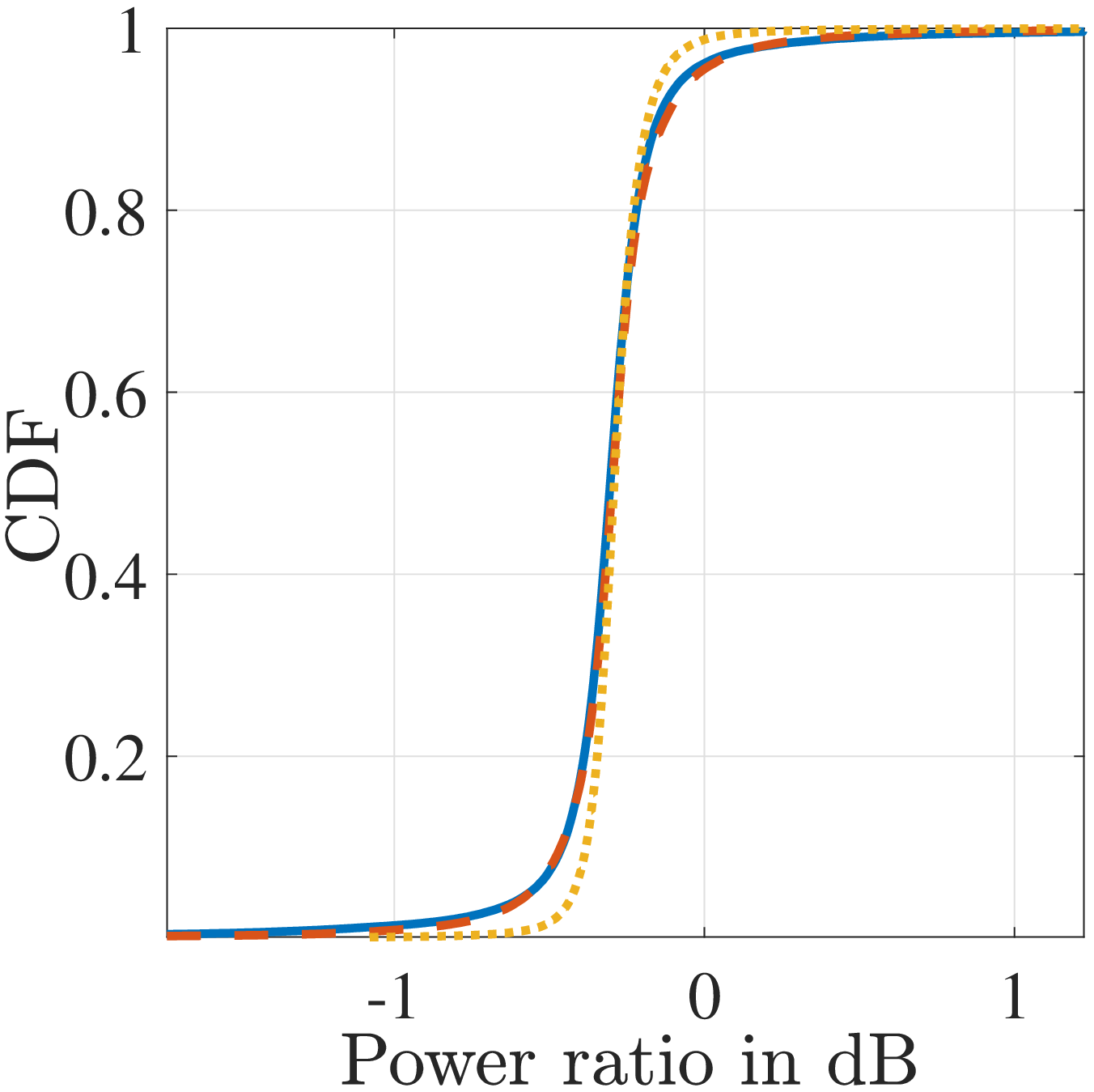}}
    \caption{(a) Empirical CDF corresponds to the ratio of computation time of direct method and the proposed algorithm and (b) Empirical CDF corresponds to the ratio of simulated power using the direct method and the proposed algorithm in dB scale for $R_{s}=N_{s}=100$ (bold blue curve), $R_{s}=N_{s}=1000$ (dashed red curve), $R_{s}=N_{s}=5000$ (dotted yellow curve)}
    \label{fig:statistics}
\end{figure}
\vspace{-0.25cm}
Firstly, we compare the complexity of the proposed simulation algorithm and the direct approach for simulating time varying vehicular wireless channels. The direct approach (used in \cite{czink2010low} for tracking channel variations outside the stationary interval) involves computation of channel parameters such as path loss and propagation delay based on the relative modification of scatterer position with respect to moving MU using geometrical calculations. We conduct simulations for varying $R_{s}$ and $N_{s}$, ranging from $R_{s}=100=N_{s}=100$ to  $R_{s}=N_{s}=5000$ (which corresponds to a model for diffuse components) in a desktop computer configured with  INTEL CORE i7\textsuperscript{\textregistered} microprocessor and $16$ GB RAM. The savings in simulation time as function of scatterer density and arrival rate is shown in Fig.~\ref{fig:statistics}(a) as an empirical cumulative distribution function (CDF) obtained from multiple realization of the scattering environment. The median of the simulation time for generating $4$s of channel using the direct approach is approximately $10$ times the simulation time for the proposed approach with $R_{s}=N_{s}=5000$. We also plot the error between the time varying received power (with unit transmit power) simulated using the direct approach and the proposed simulation algorithm. The error plot is depicted in Fig.~\ref{fig:statistics}(b) in terms of the ratio of received power simulated separately using the direct approach and the proposed simulation algorithm. The median of the error is found to be $-0.31$ dB with standard deviation equals to $0.12$ dB which can be attributed to the fact that the approximated path loss computed for the proposed simulation algorithm is less than or equal to that considered in the direct approach.
\section{Conclusion}
\label{Conclusion}
In this paper, we introduce a novel method which serves as a tool to characterize vehicular wireless channels based on the theory of ambit processes. In particular, we define suitable ambit sets and kernel functions to model scattering regions and channel properties, respectively. Key features of vehicular channels such as channel non-stationarity, multi-path death events, spatio-temporal channel correlation are studied through the proposed modeling framework. The flexibility and the tractability offered by the underlying theory of ambit processes provides a useful platform for studying important characteristics of vehicular communication channels thus paving the way for development of new channel models within the scope of next generation communication systems. Additionally, the proposed computationally efficient simulation algorithm models vehicular channels and yields several significant characteristics of vehicular communication channels. Future work involves validation of certain modeling aspects in the proposed framework with measurement data from practical scenarios.


\begin{thebibliography}{9}

\bibitem{wang2014cellular}
C.-X. Wang \textit{et al.}, ``Cellular architecture and key technologies for 5G wireless communication networks,'' \emph{IEEE Commun. Mag.}, vol. 52, no. 2, pp.~122--130, Feb. 2014.

\bibitem{Agiwal2016next}
M. Agiwal, A. Roy \textit{et al.}, ``Next generation 5G wireless
networks: A comprehensive survey,'' \emph{IEEE Commun. Surveys Tuts.}, vol. 18, no. 3, pp.~1617–1655, 3rd Quart., 2016.

\bibitem{czink2009time}
N. Czink, T. Zemen \textit{et al.}, ``A time-variant MIMO channel model directly parametrised from measurements,'' \emph{EURASIP J. Wireless Commun. Netw.}, Apr. 2009, Art. no. 687238.

\bibitem{raschkowski2015metis}
L. Raschkowski, P. Ky\"{o}sti \textit{et al.}, \emph{METIS Channel Models}, document ICT-317669-METIS/D1.4, Jul. 2015. [Online]. Available: http://www.metis2020.com

\bibitem{wu2015nonstationary}
S. Wu, C.-X. Wang \textit{et al.}, ``A non-stationary wideband channel model for massive MIMO communication systems,'' \emph{IEEE Trans. Wireless Commun.}, vol. 14, no. 3, pp. 1434–1446, Mar. 2015.

\bibitem{karedal2009geometry}
J. Karedal, F. Tufvesson \textit{et al.}, ``A Geometry-Based Stochastic MIMO Model for Vehicle-to-Vehicle Communications'', \emph{IEEE Trans. Wireless Commun.}, vol. 8, no. 7, 3646-3657, 2009

\bibitem{dahech2017nonstationary}
W. Dahech, M. P\"{a}tzold \textit{et al.}, ``A nonstationary mobile-to-mobile channel model allowing for velocity and trajectory variations of the mobile stations,'' \emph{IEEE Trans. Wireless Commun.}, vol. 16, no. 3, pp.~1987–2000, Mar. 2017.

\bibitem{patzold2012mobile}
M. P\"{a}tzold, \emph{Mobile Radio Channels}, 2nd ed., U.K.: Wiley, 2012

\bibitem{wang2018survey}
C.-X. Wang, Ji Bian \textit{et al.}, ``A survey of 5G channel measurements and models", \emph{IEEE Commun. Surveys Tuts.}, vol. 20, no. 4, pp.~3142-3168, 4th Quart., 2018.

 \bibitem{zajic2006efficient}
 A. G. Zaji\'{c} and G. L. St\"{u}ber, ``Efficient simulation of Rayleigh fading with enhanced de-correlation properties,” \emph{IEEE Trans. Wireless Commun.}, vol. 5, no. 7, pp.~1866--1875, 2006.
  
 \bibitem{Kaltenberger2007low}
 F. Kaltenberger, T. Zemen, and C. W. Ueberhuber, ``Low-complexity geometry-based MIMO channel simulation,'' \emph{EURASIP Journal on Advances in Signal Processing} no. 1, 2007
\bibitem{czink2010low}  
N. Czink, F. Kaltenberger \textit{et al}, ``Low-complexity geometry-based modeling of diffuse scattering." \emph{In Proceedings of the fourth European conference on antennas and propagation}, pp. 1-4. IEEE, 2010.

 \bibitem{barndorff2018ambit}
 O. E. Barndorff-Nielsen, F. E. Benth \textit{et al.}, ``Ambit Stochastics'', Springer, New York, vol. 88, 2018.
  
 \bibitem{Ertel1999angle}
 R. B. Ertel and J. H. Reed, ``Angle and Time of Arrival Statistics for Circular and Elliptical Scattering Models'', \emph{IEEE J. Sel. Areas Commun.}, vol. 17, no. 11, pp.~1829--1840, 1999

\bibitem{bennedsen2014discretization}
M. Bennedsen, L. Asger \textit{et al.}, ``Discretization of L\'{e}vy semistationary processes with application to estimation," arXiv preprint arXiv:1407.2754, 2014.

\end{thebibliography}
\end{document}